\documentclass[english]{article}
\usepackage{graphicx,color}
\def\no{\noindent}
\def\bc{\begin{center}}
\def\ec{\end{center}}

\def\beq{\begin{equation}}
\def\eeq{\end{equation}}

\def\br{{\bf r}}
\def\bx{{\bf x}}
\def\bR{{\bf R}}

\def\bk{{\bf k}}

\def\be{{\bf e}}

\begin{document}

\title{
An invariant measure of chiral quantum transport 
}

\author{K. Ziegler$^*$\\
Institut f\"ur Physik, Universit\"at Augsburg\\
D-86135 Augsburg, Germany\\
$^*$email: klaus.ziegler@physik.uni-augsburg.de
}

\maketitle

\no
Abstract:

We study the invariant measure of the transport correlator for a chiral Hamiltonian and analyze its 
properties. The Jacobian of the invariant measure is a function of random phases.
Then we distinguish the invariant measure before and after the phase integration.
In the former case we found quantum diffusion of fermions and a uniform zero mode that
is associated with particle conservation. After the phase integration the transport correlator
reveals two types of evolution processes, namely classical diffusion and back-folded random walks.
Which one dominates the other depends on the details of the underlying chiral Hamiltonian
and may lead either to classical diffusion or to the suppression of diffusion.

\section{Introduction}
\label{sect:introduction}

We consider the quantum evolution of a particle from the site $\br'$ to the site $\br$ during the time period $t$ 
in a particle conserving system.
Then the building block for quantum transport is provided by the transition probability $|\langle\br|e^{-iHt}|\br'\rangle|^2$,
which is obtained from the transition amplitude $\langle\br|e^{-iHt}|\br'\rangle$
over time $t$ for a system defined by the Hamiltonian $H$. We can Laplace transform the amplitude with respect to 
time and calculate the resulting transition probability, the transport correlator, as
\beq
\label{transport_c}
\Big|\int_0^\infty\langle \br|e^{-iHt}|\br'\rangle e^{-\epsilon t} dt\Big|^2
=|G^{}_{\br\br'}(\epsilon)|^2
,
\eeq
where $G(\epsilon)=(H-i\epsilon)^{-1}$ is the one-particle Green's function. Besides the spatial coordinates 
$\br$, the Hamiltonian can also depend on additional quantum numbers, such as a spinor or band index. Quantum 
transport in multiband systems has been a popular subject in recent years due to the discovery of new materials,
beginning with graphene~\cite{no.ge.mo.ji.ka.gr.du.fi.05}, 
topological insulators~\cite{PhysRevB.78.195424,bernevig2013topological,PhysRevLett.123.046801} 
and including more specialized models such as two-dimensional models with random 
gauge field~\cite{PhysRevB.106.L081410}.

To describe realistic transport in a disordered material we must include random scattering.
This can be attributed to a random Hamiltonian $H$.
A very successful approach to random scattering is based on the 
random matrix theory~\cite{wigner1955,dyso.62,me.04,Edelman_Rao_2005,Amir_2020},
where $H$ is a real symmetric or Hermitian matrix with independently and identically distributed matrix 
elements $H_{ij}$ for $i\le j$.
The eigenvalues of a real symmetric matrix are invariant under an orthogonal transformation, 
which implies $H=O^T D O$ with the diagonal matrix $D$ that comprises the eigenvalues of $H$.
Then the Jacobian of the transformation from the distribution of $H$
to the distribution of the diagonal matrix $D$ and the orthogonal matrices $O$ characterizes the
orthogonal random matrix ensemble. Other ensembles (e.g., Hermitian matrices) can be distinguished 
according to their Jacobian with respect to the symmetry transformation.
Many interesting properties (e.g., level repulsion statistics) can be calculated from the invariant 
measure~\cite{me.04,Edelman_Rao_2005,Amir_2020}. This is based on the idea that the fluctuations of the
$N$ eigenvalues of, for instance, a symmetric $N\times N$ random matrix $H$, 
which are invariant under the orthogonal transformation of the matrix, are separated from the remaining
$N(N-1)/2$ random degrees of freedom of the matrix elements. Then the Jacobian of the latter provides the invariant 
measure. This fundamental concept can also be applied to a more general class of random matrices, 
such as the real symmetric matrices formed by the elements 
$|G^{}_{\bx\bx'}|^2\equiv G^{}_{\bx\bx'}G^\dagger_{\bx'\bx}$, where $\bx=(\br,\mu)$ represents the
lattice site $\br$ and the spinor or band index $\mu=1,2$ of the chiral structure.
In analogy to the symmetric matrix ensemble mentioned above, we must identify the symmetry, which depends 
on the underlying Hamiltonian of the Green's function.
This was studied in Ref.~\cite{zieg.15} for a lattice hopping Hamiltonian 
with chiral symmetry and will be briefly discussed in Sect. \ref{sect:model}.
Instead of fixing the eigenvalues of these matrices, one can also use a uniform 
approximation of the matrix that is fixed by a variational approach (saddle-point approximation). Then the 
fluctuations, induced by the chiral symmetry, are represented as the invariant measure. 

The goal of the following analysis is to start from the disorder averaged transport correlator 
$\langle |G^{}_{\bx\bx'}(\epsilon)|^2\rangle_d$ and reduce the general disorder average 
$\langle \ldots\rangle_d$ by the integration with respect to its invariant measure. 
The justification of this reduction is that the long-range behavior is associated with the invariant measure 
of the underlying chiral property of a particle-hole symmetric Hamiltonian $H$. It will be shown that
the invariant measure provides the long-range properties even before performing the phase integration.

The paper is organized as follows. After a brief discussion of the chiral 
Hamiltonian $H$, the related Green's functions, and the corresponding chiral invariance in Sect. \ref{sect:model}, 
we give a summary of the invariant measure, which was originally derived in a previous paper, in Sect. \ref{sect:IM}.
This is followed by an analysis of the corresponding Jacobian and a discussion of the effects caused by
phase averaging. In Sect. \ref{sect:discussion} the different mappings, related to the derivation of the invariant
measure, are summarized and an example for a simple two-band Hamiltonian is studied. Finally, we provide
some conclusions with open problems and a brief outlook in Sect. \ref{sect:conclusions}.

\section{Model}
\label{sect:model}

The transport correlator in Eq. (\ref{transport_c}) is a fundamental quantity for the description of 
quantum transport~\cite{RevModPhys.83.407}, 
consisting of products of matrix elements of the advanced and the retarded one-particle Green's function 
$G(\pm\epsilon)=(H\pm i\epsilon)^{-1}$ in the form of $G_{\bx\bx'}(\epsilon)G_{\bx'\bx}(-\epsilon)
=(H- i\epsilon)^{-1}_{\bx\bx'}(H+ i\epsilon)^{-1}_{\bx'\bx}$. 
In the following we consider a multiband Hamiltonian $H$ on a $d$-dimensional lattice $\Lambda$ 
with $|\Lambda|<\infty$ sites and assume that it is 
related to its transposed $H^T$ through the chiral relation $UH^TU^\dagger=-H$, where
$U$ is a spatially uniform unitary matrix. Then the chiral relation implies that the trace and the determinant
for the advanced/retarded Green's functions obey ${\rm Tr}[G(-\epsilon)]=-{\rm Tr}[G(\epsilon)]$ and 
$\det[G(-\epsilon)]=\det[-G(\epsilon)]$. Using the Pauli matrices $\{\sigma_j\}_{j=1,2,3}$ and
the $2\times 2$ unit matrix $\sigma_0$, a two-band Hamiltonian can be expressed by an expansion in terms
of these matrices. An example is the Hamiltonian $H=h_1\sigma_1+h_2\sigma_2 +m\sigma_3$
with a random $m$ that is independently and identically distributed on the lattice, while the matrices $h_{1,2}$
act on the lattice. For symmetric matrices $h_{1,2}$
this Hamiltonian satisfies the chiral relation for $U=\sigma_2$.

The pair of the advanced and retarded Green's functions can also be written as blocks of an advanced 
Green's function when we replace the Hamiltonian $H$ by the block-diagonal Hamiltonian ${\hat H}={\rm diag}(H,H^T)$. 
Then the corresponding advanced Green's function reads ${\hat G}(\epsilon)=({\hat H}-i\epsilon)^{-1}$
and we get $G(\epsilon)={\hat G}_{11}(\epsilon)$ and $G(-\epsilon)=-U{\hat G}_{22}(\epsilon)U^\dagger$.
For
\beq
\label{generator}
{\hat S}=\pmatrix{
0 & sU \cr
s'U^\dagger & 0 \cr
}
\eeq
with two independent parameters $s$ and $s'$ we obtain for ${\hat H}$ the symmetry relation
\beq
\label{symm00}
e^{\hat S}{\hat H}e^{\hat S}={\hat H}
,
\eeq
since ${\hat H}$ and ${\hat S}$ anticommute. It was shown that this symmetry is associated with an invariant 
measure~\cite{zieg.15}. 

\section{Invariant measure}
\label{sect:IM}

In general, for a random chiral Hamiltonian there exists an invariant measure of the disorder average. 
To demonstrate this we consider the example of the previous section $H=h_1\sigma_1+h_2\sigma_2 +m\sigma_3$
and assume a random $m_\br$ at each lattice site $\br$ that obeys a Gaussian distribution $\exp(-m_\br^2/g)dm_\br$. 
To control the strong fluctuations of $|{\hat G}_{\bR\bR'}(\epsilon)|^2$ near the poles of the Green's functions,
we map the random field $m_\br$ to a random $4\times4$ matrix field $Q_\br$, where the latter
reflects the matrix structure of ${\hat G}$ in Sect. \ref{sect:model}. This mapping provides the relation 
\beq
\langle |{\hat G}_{\bR\bR'}(\epsilon)|^2\rangle_d=\langle Q_\bR Q_{\bR'}\rangle_Q
.
\eeq
Thus, the average transport correlator is a correlation function of the random field 
$Q_\br$~\cite{PhysRevB.79.195424}. The mapping $m_\br\to Q_\br$
yields a Jacobian, and the Gaussian weight $\exp(-m_\br^2/g)$ transforms to the corresponding weight 
$P(\{Q_\br\})$ for the field $Q_\br$, where the latter is correlated on the lattice.

Next, we include the symmetry transformation of Eq. (\ref{symm00}) by writing $Q=e^{\hat S} Q_0e^{\hat S}$, 
where $Q_0$ represents all degrees of freedom of $Q$ which are not related to the symmetry transformation.
${\hat S}$ of Eq. (\ref{generator}) is now the space-dependent generator of the symmetry transformation after
replacing the parameters $s$ and $s'$ by a field $\varphi_\br$ and its conjugate $\varphi'_\br$. Then we
approximate $Q_0$
through a saddle-point approximation and  keep only the Jacobian with respect to the fields $\varphi_\br$, $\varphi'_\br$ 
as relevant integration variables because they provide the generators of the symmetry transformation. In other words,
the integration over $Q_\br$ is separated into two parts, namely one that leaves $P(\{Q_\br\})$ 
invariant and one that does not. The former, together with the corresponding Jacobian $J$, 
defines an invariant measure. In a final step we map the $\varphi$, $\varphi'$ integration to an integration
over a complex two-component vector field. It turns out that the integration over the modulus of the 
field components can be performed and we remain with a field that consists of two components 
$(\exp(i\alpha_{\br1}),  \exp(i\alpha_{\br2}))$. As a subtle point it should be noted that, in contrast to the random matrix
approach, there is a symmetry breaking term in ${\hat G}(\epsilon)$ due to $\epsilon$. Its role for the creation of
a zero mode in the limit $\epsilon\to0$ was discussed in detail in previous 
works~\cite{PhysRevB.79.195424,zieg.15,Ziegler_2022}.
More specific, it was shown that for a chiral Hamiltonian the 
invariant measure is given by an integration over random phases, where the Jacobian is $J=\det C$ with  
the $|\Lambda|\times|\Lambda|$ random-phase matrix $C$. The elements of this matrix are~\cite{zieg.15}
\beq
\label{C_matrix00}
C_{\br\br'}=
2\delta_{\br\br'}-\sum_{\mu,\mu'=1,2}z_{\br\mu} h_{\br\mu,\br'\mu'}
\sum_{\br''}\sum_{\mu''=1,2}h^\dagger_{\br'\mu',\br''\mu''}z^*_{\br''\mu''}
\ {\rm with}\ z_{\br\mu}=e^{i\alpha_{\br\mu}} \ (0\le \alpha_{\br\mu}<2\pi)
\eeq
with the non-random $2|\Lambda|\times2|\Lambda|$ matrix
\beq
\label{eff_green}
h={\bf 1}+2i\eta(\bar{H}-i\eta-i\epsilon)^{-1}
,
\eeq
which represents an effective Green's function.
$\eta$ ($0<\eta<\infty$) is an effective scattering rate, usually obtained either from experimental
observations, from numerical simulations or from a self-consistent Born approximation
of the average one-particle Green's function. It is proportional to the average density of states,
indicating that on average there are state with non-zero density for $\eta>0$. 
For our approach it must be positive but its actual value is not important for the subsequent discussion. 
Finally, ${\bar H}$ is the average Hamiltonian: ${\bar H}=\langle H\rangle_d$.
It is crucial to note that in the limit $\epsilon\to 0$ the matrix $h$ is unitary (i.e., $hh^\dagger={\bf 1}$).
$\epsilon>0$ is only necessary to avoid the uniform zero mode with $\sum_{\br'}C_{\br\br'}=0$.
After removing it from the spectrum (cf. App. \ref{app:zero_mode1}), we can take the limit $\epsilon\to0$.
The phase integration provides the relation
\beq
\label{corr0}
\langle |G^{}_{\bR\bR'}(\epsilon)|^2\rangle_{d}
\sim\frac{\langle {\rm adj}_{\bR\bR'} C\rangle_\alpha}{\langle\det C\rangle_\alpha}
,
\eeq
where the adjugate matrix is ${\rm adj}_{\bR\bR'} C\equiv C^{-1}_{\bR\bR'}\det C $, which is obtained from the
determinant by differentation as 
\beq
\label{adjugate1}
\partial_{W_{\bR'\bR}}\det (C+W)|_{W=0}
=\sum_\pi \delta_{\bR\pi(\bR')}{\rm sgn}\pi \prod_{\br\ne\bR'}C_{\br\pi(\br)}
.
\eeq
The relation $\sim$ preserves the asymptotic properties on large distances $|\bR-\bR'|$. 
The phase average $\langle \ldots\rangle_\alpha$ is taken with respect to a statistically independent and uniform
distribution of the phases $\{\alpha_{\br\mu}\}$ on the interval $[0,2\pi)$. 
We will discuss the phase average only in a second step but treat the invariant measure in the following 
for a general realization of the random phases.
Averaging can be performed explicitly and leads to a loop expansion of the determinant and the 
adjugate matrix~\cite{zieg.15}.

The random-phase matrix $C$ has some remarkable properties, which will be analyzed subsequently, revealing eventually
a spatial scaling behavior of $\langle |G^{}_{\bR\bR'}(\epsilon)|^2\rangle_{d}$ and quantum diffusion
of fermions. To this end we consider the random unitary matrix $u=zhz^*$ and rewrite $C$ as
\beq
\label{C_matrix0}
C_{\br\br'}={\rm Tr}_2[{\hat C}_{\br\br'}E_2]
\ \ \  {\rm with}\ \ 
{\hat C}_{\br\mu,\br'\mu'}
=\delta_{\br\br'}\delta_{\mu\mu'}-u_{\br\mu,\br'\mu'}\sum_{\br'',\mu''}u^\dagger_{\br'\mu',\br''\mu''}
,
\eeq
where $E_2=\sigma_0+\sigma_1$ and ${\rm Tr}_2$ is the trace with respect to $2\times2$ matrices for a Hamiltonian 
$H$ with two bands. A generalization to a multiband Hamiltonian is straightforward but not considered explicitly.
Moreover, with the $2|\Lambda|\times 2|\Lambda|$ matrix $E$, whose elements are 1, we can write
$[u^\dagger E]_{\br\mu}=\sum_{\br'',\mu''}u^\dagger_{\br\mu,\br''\mu''}$ and get
\beq
\label{C_matrix1}
{\hat C}={\bf 1}-u[u^\dagger E]_d
={\bf 1}-uu^\dagger +u(u^\dagger-[u^\dagger E]_d)
,
\eeq
where $[A]_d$ is the diagonal part of the general matrix $A$ and 
${\bf 1}-uu^\dagger=4\epsilon\eta[{\bar H}^2+(\eta+\epsilon)^2]^{-1}$.
Now we split $u^\dagger$ into its diagonal part $u':=[u^\dagger]_d$ and its off-diagonal part 
$v:=u^\dagger-u'$ and insert this into $u^\dagger-[u^\dagger E]_d$. This gives
\beq
\label{off-diag_part}
u'+v-[u'E]_d-[vE]_d=v-[vE]_d
,
\eeq
since $[u'E]_d=u'$. Thus, the diagonal part $u'$ drops out and reveals that 
$z^*(v-[vE]_d)z=h^\dagger-[h^\dagger]_d-[vE]_d$ can be understood as a hopping matrix $z^*vz$ with an 
additional random diagonal part $-[vE]_d$. In contrast to the conventional disordered tight-binding 
model~\cite{ande.58,ab.an.79,wegner79,PhysRevB.22.4666} 
(an overview of the more recent progress in this field can be found in 
Refs.~\cite{doi:10.1142/7663,stolz2011introduction}) the diagonal term also scales 
with the hopping matrix $v$, since it consists of a sum of hopping terms. In other words, when $v$
changes as $v\to L^p v$ under a change of the lattice constant $a\to La$,
the entire matrix $v-[vE]_d$ scales with $L^p$. 

We return to $C$ in Eq. (\ref{C_matrix0}) and ${\hat C}$ in Eq. (\ref{C_matrix1}) and write
\beq
\label{final1}
C=\epsilon C'+C''
\ \ {\rm with}\  \ 
C'=4\eta{\rm Tr}_2[(z{\bar H}^2z^*+(\eta+\epsilon)^2)^{-1}E_2]
, \ \ 
C''={\rm Tr}_2[u(v-[vE]_d)E_2] 
.
\eeq
The properties of the determinant and the adjugate matrix of $C$ are linked to the spectral properties of 
${\bf 1}-{u^\dagger}^{-1}[u^\dagger E]_d$ in Eq. (\ref{C_matrix1}). To analyze this matrix we rewrite
it as a quadratic form in $z$:
\beq
\label{QF}
z_{\br\mu}(\delta_{\br\br'}\delta_{\br\br''}\delta_{\mu\mu''}
-\sum_{\mu'}h^{\dagger -1}_{\br\mu,\br'\mu'}h^\dagger_{\br'\mu',\br''\mu''})z^*_{\br''\mu''}
\]
\[
=z_{\br\mu}\sum_{\mu'}(\delta_{\br\br'}
\sum_{\bar{\br}}h^{\dagger -1}_{\br\mu,\bar{\br}\mu'}h^\dagger_{\bar{\br}\mu',\br''\mu''}
-h^{\dagger -1}_{\br\mu,\br'\mu'}h^\dagger_{\br'\mu',\br''\mu''})z^*_{\br''\mu''}
,
\eeq
which yields
\beq
\label{q_diffusion}
C''_{\br\br'}=2\sum_{\bar{\br}}(\delta_{\br\br'}-\delta_{\bar{\br}\br'})K_{\br\bar{\br}}
=2(\delta_{\br\br'}\sum_{\bar{\br}}K_{\br\bar{\br}}-K_{\br\br'})
,
\eeq
where
\beq
\label{kernel}
K_{\br\br'}
=\frac{1}{2}\sum_{\mu,\mu',\mu''}\sum_{\br''}
u^{\dagger -1}_{\br\mu,\br'\mu'}u^\dagger_{\br'\mu',\br''\mu''}
\ \ {\rm with}\ \
\sum_{\bar{\br}}K_{\br\bar{\br}}=1
.
\eeq
This indicates that $K_{\br\br'}$ is formally a random transition amplitude
for $\br'\to\br$ with particle conservation due to $\sum_{\bar{\br}}K_{\br\bar{\br}}=1$.
Then $C''$, defined in Eq. (\ref{q_diffusion}), can be understood as a model for quantum diffusion. 
In contrast to classical diffusion its transition rates are complex numbers rather than 
probabilities. On the other hand, the average amplitude is a positive number, given by 
the isotropic mean 
\beq
\label{mean_K}
\langle K_{\br\br'}\rangle_\alpha
=\frac{1}{2}\sum_{\mu,\mu'}h^{\dagger -1}_{\br\mu,\br'\mu'}h^\dagger_{\br'\mu',\br\mu}
=\frac{1}{2}\sum_{\mu,\mu'}|h^{}_{\br\mu,\br'\mu'}|^2
\eeq
for $h^{\dagger -1}=h$ in the limit $\epsilon\to0$. This implies
\beq
\label{mean_C''}
\langle C''_{\br\br'}\rangle_\alpha
=2(\delta_{\br\br'}-\langle K_{\br\br'}\rangle_\alpha)
=2(\delta_{\br\br'}-\sum_{\mu,\mu'}|h^{}_{\br\mu,\br'\mu'}|^2)
.
\eeq

These results can be interpreted as follows. Before phase average
the expansion of the determinant $\det C$ or the adjugate matrix ${\rm adj}_{\bR\bR'} C$
provide quantum walks on the lattice in the form of loops, where a site is only visited
once. This reflects the fact that the determinant represents fermions, which obey
Pauli's exclusion principle. Thus, the invariant measure describes quantum diffusion
of fermions.

\begin{figure}[t]
    \centering
\includegraphics[width=9cm,height=3cm]{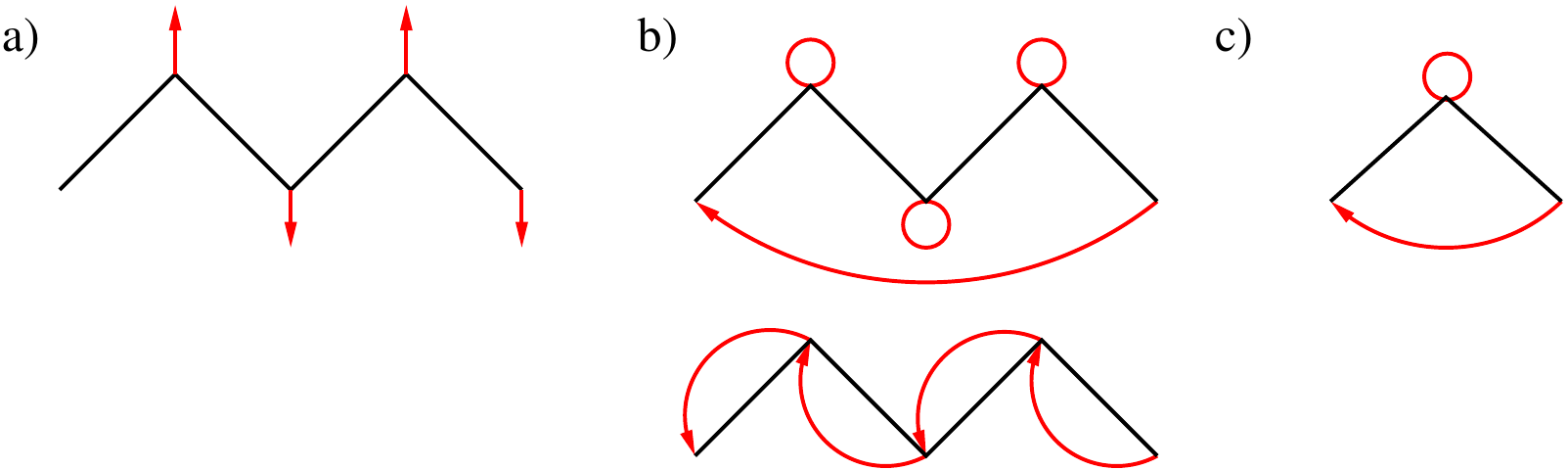}
    \caption{a) A realization of a quantum walk from site $\br_1$ to site $\br_5$ before phase
    average, consisting     of four steps: $K_{\br_1\br_2}K_{\br_2\br_3}K_{\br_3\br_4}K_{\br_4\br_5}$. 
    According to Eq. (\ref{kernel}), the black links represent $u^{\dagger -1}$, 
    while the red arrows represent the factors $\sum_{\br'',\mu''}u^\dagger_{\br\mu,\br''\mu''}$.
    b) The same random walk after phase average: Two typical contributions are shown for
    $\langle K_{\br_1\br_2}K_{\br_2\br_3}K_{\br_3\br_4}K_{\br_4\br_5}\rangle_\alpha$,
    where the red connection is $h^\dagger$ and corresponds to a ``rubber band''. The lower graph represents
    a classical random walk $\langle K_{\br_1\br_2}\rangle_\alpha\langle K_{\br_2\br_3}
    \rangle_\alpha\langle K_{\br_3\br_4}\rangle_\alpha\langle K_{\br_4\br_5}\rangle_\alpha$.
    c) This is the two step walk $\langle K_{\br\br'}K_{\br'\br''}\rangle_\alpha
    -\langle K_{\br\br'}\rangle_\alpha\langle K_{\br'\br''}\rangle_\alpha$ of Eq. (\ref{two_steps}).
    }    
\label{fig:graph1}
\end{figure}

After the phase average the physics is quite different though, since the average
connects lattice sites $\br$ and $\br'$, visited by the quantum walk,
with the transition amplitude $h^\dagger_{\br\br'}$. The latter decays on large distances
exponentially, while on short distances it depends strongly on the specific structure of
the Hamiltonian ${\bar H}$.
This is as if we connect the quantum walk between the two lattice sites by an elastic rubber 
band with an exponential strength for large stretches.
In other words, phase averaging is subject to quantum interference effects, which is sensitive
to spatial directions and can cause an anisotropic evolution. The average transition amplitude
in Eq. (\ref{mean_K}) is isotropic, since $\langle K_{\br\br'}\rangle_\alpha$ is equal in all 
lattice directions. Therefore, the length of the rubber band is the same in all directions. 
The situation is different though for two consecutive steps $\br\to\br'\to\br''$:
\beq
\label{two_steps}
\langle K_{\br\br'}K_{\br'\br''}\rangle_\alpha
=\langle K_{\br\br'}\rangle_\alpha\langle K_{\br'\br''}\rangle_\alpha
+\frac{1}{4}\sum_{\mu,\mu',\mu'',\mu'''}
h^{\dagger -1}_{\br\mu,\br'\mu'}h^\dagger_{\br'\mu',\br'\mu'''}
h^{\dagger -1}_{\br'\mu''',\br''\mu''}h^\dagger_{\br''\mu'',\br\mu}
.
\eeq
While the first term on the right-hand side is isotropic, the second term depends on the
site $\br''$ relative to $\br$ and $\br'$. This is visualized in Fig. \ref{fig:graph1}c)
and calculated for a special example in Sect. \ref{sect:discussion}.
While for the first term the length of the rubber band does not change when going from $\br'$ to any
of the available $\br''$ with fixed length $|\br''-\br'|$, the rubber band has different lengths 
for the second term. The latter favors a return of the quantum walk closer to $\br$ rather than to 
move away from $\br$. This picture can be generalized to more steps, connected by rubber bands between
pairs of sites along the quantum walk. 
This can be illustrated by a graphical representation in terms of loops and a string:

Before averaging the determinant is graphically a distribution of loops on the lattice, formed by connections
$h_{\br\mu,\br'\mu'}$, where a lattice site is visited only once. The loops consist of $l$ connections, 
including $l=0$ for a single isolated site. 
This structure originates in the determinant, which reads $\det C=\sum_\pi{\rm sgn}\pi\prod_{\br}C_{\br\pi(\br)}$.
Upon averaging the factor
$\chi_{\br'\mu}:=\sum_{\br''}\sum_{\mu''=1,2}h^\dagger_{\br'\mu',\br''\mu''}z^*_{\br''\mu''}$
in $C_{\br\br'}$, is crucial since its $z^*$ factor compensates a $z$ factor of the loops and, consequently, connects sites 
$\br'$ and $\br''$. Graphically, this creates
pairwise connections by $h^\dagger_{\br'\mu',\br''\mu''}$ on lattice sites $\br'$ and $\br''$, such that the loops,
comprising of connections $h$, are linked by $h^\dagger$. The latter are the rubber bands, and the general graphs
consist of 4-vertices that are connected to each other by two $h$ and by two $h^\dagger$.
In other words, we have quantum walks consisting of $h$ connections and $h^\dagger$ connections. 
The lattice is partially covered by these quantum walks, while the remaining sites are isolated. The latter are
created by the diagonal elements $C_{\br\br}$, which are reduced to the $\br''=\br$, $\mu''=\mu$ term of the 
internal sum, such that their weight is $2-\sum_{\mu,\mu'=1,2}h_{\br\mu,\br\mu'}h^\dagger_{\br\mu',\br\mu}$.
The adjugate matrix $\langle {\rm adj}_{\bR\bR'} C\rangle_\alpha$ is obtained by differentation of the 
determinant (cf. Eq. (\ref{adjugate1})). This means graphically that we cut a loop that contains the sites $\bR$ and $\bR'$ 
and create a string of $h$'s connecting these two sites.  The endpoints of the string are 2-vertices, as illustrated in
Figs. \ref{fig:graph1}b, c.
There is a specific pairing of sites, where the lengths of the rubber bands are not affected for all possible
conformations of the random walks, namely when the sites of the loops and the string are pairwise 
connected by a pair of $h$ and $h^\dagger$. This corresponds to the factorization of the phase average
\[
\langle K_{\br\br_1}\rangle_\alpha\cdots\langle K_{\br_{n-1}\br_n}\rangle_\alpha
\ ,
\]
which is visualized in the lower graph of Fig. \ref{fig:graph1}b). It represents a classical random walk
with transition probabilities $\langle K_{\br\br'}\rangle_\alpha\ge 0$ and
$\sum_{\bar{\br}}\langle K_{\br\bar{\br}}\rangle_\alpha=1$. The co-existence of the
classical random walk and the folded walks with rubber bands indicates
a competition between those walks. To understand this point better we rewrite
the right-hand side of Eq. (\ref{corr0}) as
\beq
\label{expansion}
\frac{\langle {\rm adj}_{\bR\bR'}C\rangle_\alpha}{\langle\det C\rangle_\alpha}
=\frac{{\rm adj}_{\bR\bR'}\langle C\rangle_\alpha+\sum_j a_{j;\bR\bR'}}{\det\langle C\rangle_\alpha +\sum_j d_j}
,
\eeq
where the two sums are finite on a finite lattice and are
created by a systematic expansion in terms of truncated correlations (linked cluster expansion) as
\beq
\label{truncated_expansion}
\langle\prod_\br C_{\br\pi(\br)}\rangle
=\prod_\br\langle C_{\br\pi(\br)}\rangle+\sum_{\br,\br'}
\left[\langle C_{\br\pi(\br)} C_{\br'\pi(\br')}\rangle_\alpha-\langle C_{\br\pi(\br)}\rangle_\alpha
\langle C_{\br'\pi(\br')}\rangle_\alpha\right]
\prod_{\bar{\br}\ne \br,\br'}\langle C_{\bar{\br}\pi(\bar{\br})}\rangle_\alpha +\cdots
\eeq
for the determinant and the adjugate matrix. 
In terms of graphs the expansions in Eq. (\ref{expansion}) consist of loops (loops and a string) for the denominator
(numerator), which are connected by 4-vertices. Then it depends on the 
details of the Hamiltonian ${\bar H}$ and on the scattering rate $\eta$, which term 
dominates the others. On an elementary level, we can compare the first and the second
term on the right-hand side of Eq. (\ref{two_steps}), as discussed in Sect. \ref{sect:discussion}. 
For large distances the classical random walk would be important. On the other
hand, for a given number of steps the sum over all realizations of the classical random walk
may have a much smaller weight than the sum over all compactly folded walks, which are 
created by the rubber bands. Thus, we must compare the weights of all walks with a given 
number of steps and extract those with the highest weight as dominant. 

\section{Discussion of the Results}
\label{sect:discussion}

This work on the invariant measure of the transport properties in chiral
systems is based on the random phase matrix $C=\epsilon C'+C''$ of Eq. (\ref{final1}),
whose properties were studied in the previous section.
$C$ is connected with the original random Hamiltonian $H$ through 
several mappings between different random matrices. First, there is the mapping between Hamiltonians
\beq
H\to z{\bar H}z^*-i\eta
\eeq
with the phenomenological scattering rate $\eta>0$.
This is accompanied by the mapping of the Green's function $(H-i\epsilon)^{-1}\to u$ with
the effective Green's function of the Hamiltonian $z{\bar H}z^*-i\eta$
\beq
\label{eff_gf2}
u=zhz^*={\bf 1}+2i\eta (z{\bar H}z^*-i\eta-i\epsilon)^{-1}
,
\eeq
from which we eventually get $C$ in Eq. (\ref{final1}) and the transition amplitude $K$ 
in Eq. (\ref{kernel}). $u$ is unitary in the limit $\epsilon\to0$,
which implies the conservative transfer $\sum_{\br'}K_{\br\br'}=1$ in Eq. (\ref{q_diffusion}).
The eigenvalues of $u$ are identical with those of $h$ in Eq. (\ref{eff_green}), which is not random. 
The corresponding eigenvectors $\Psi_{\bk,\br\mu}$ of $u$ are random though, 
given by the plane wave $\psi_{\bk,\br}=\exp(i\bk\cdot\br)/\sqrt{|\Lambda|}$ with a random phase factor
as $\Psi_{\bk,\br\mu}=z_{\br\mu}\psi_{\bk,\br}$.

For strong scattering ($\eta\sim\infty$) we get $u\sim -{\bf 1}$, whereas for weak scattering ($\eta\sim0$)
we obtain $u\sim {\bf 1}$. In both limits the transition matrix becomes diagonal with $K={\bf 1}$, 
implying that the quantum
diffusion disappears. All these results are valid before phase averaging. The transport correlator,
defined through the relation in Eq. (\ref{corr0}), requires phase averaging of $\det C$ and the related
adjugate matrix. Although the average can be performed easily, the result leads to complex expressions,
which can be associated with classical random walks and folded random walks with rubber bands.
On a more elementary level we get for $u$ the diagonal mean 
$\langle u_{\br\mu,\br'\mu'}\rangle_\alpha=\delta_{\br\br'}\delta_{\mu\mu'}h_{\br\mu,\br\mu}$ and 
a vanishing variance, while the mean $\langle C''\rangle_\alpha$ in Eq. (\ref{mean_C''}) reveals classical 
diffusion. The fluctuations of $C''$ are characterized by the variance
\beq
\label{variance}
\langle {C''_{\br\br'}}^2\rangle_\alpha-\langle C''_{\br\br'}\rangle_\alpha^2
=\sum_{\mu}\left(\sum_{\mu'}|h_{\br\mu,\br'\mu'}|^2\right)^2
.
\eeq
Then the ratio of the standard deviation and the mean value is
\beq
\label{ratio_R}
R:=\frac{\sqrt{\langle C_{\br\br'}^2\rangle_\alpha-\langle C_{\br\br'}\rangle_\alpha^2}}
{\langle C_{\br\br'}\rangle_\alpha}
=\sqrt{1-2p_1p_2/(p_1+p_2)^2}
\ \ {\rm with}\ \ p_\mu=\sum_{\mu'}|h_{\br\mu,\br'\mu'}|^2
,
\eeq
which is equal or less than 1. In other words, the fluctuations of $C$ around the mean value $\langle C\rangle$
are restricted. Moreover, for $N$ bands with equal $p_\mu\equiv {\bar p}$ we get $R=N^{-1/2}$, indicating a 
non-random limit for a large number of bands $N\sim\infty$. In contrast, the fluctuations of 
$|G_{\br\br'}(\epsilon)|^2$ are unbounded for $\epsilon\sim0$ due to the proximity to the
poles of the Green's functions. This demonstrates that the invariant measure is better controlled
than the original transport correlator. 

In order to give an example for the anisotropic effect of the phase averaging,
we expand $h$ in powers of $1/\eta$:
\beq
h=-{\bf 1}+2i\frac{1}{\eta}{\bar H}+\frac{2}{\eta^2}{\bar H}^2+O(1/\eta^3)
.
\eeq
Then we choose a band-diagonal matrix ${\bar H}=\Delta\sigma_3$, where $\Delta$ is 
nearest-neighbor hopping:
$
\Delta_{\br-\br'}=1
$
for $\br$, $\br'$ nearest-neighbor sites and zero otherwise. For the nearest-neighbor 
vectors $\br'=\br\pm \be_{1,2}$ on a square lattice with lattice unit vectors $\be_{1,2}$
we obtain, after neglecting terms of order $1/\eta^3$, 
\beq
\label{nn_hopping}
h_{\br,\mu}=\cases{
-1+8/\eta^2 & $\br=0$ \cr
2(-1)^\mu i/\eta & $\br=\pm \be_{1,2}$ \cr
2/\eta^2 & $\br=\pm(\be_1+\be_2),\pm(\be_1-\be_2),\pm2\be_{1,2}$  \cr
0 & otherwise \cr
}
\eeq
with $h_{\br\mu,\br'\mu}\equiv h_{\br-\br',\mu}$.
The weight $|h_{\br\mu}|$ decreases in powers of $1/\eta$ as we increase the distance $|\br|$,
with the maximal weight $1-8/\eta^2$ for $\br=0$.
This yields for the classical diffusion in two steps with the special choice 
$\br'=\br+\be_1$ and $\br''=\br'+\bR=\br+\be_1+\bR$ 
\beq
\langle K_{\br\br'}\rangle_\alpha\langle K_{\br'\br''}\rangle_\alpha
=\frac{1}{4}\sum_{\mu}|h_{\br-\br',\mu}|^2\sum_{\mu'}|h_{\br'-\br'',\mu'}|^2
=\frac{1}{4}\sum_{\mu}|h_{-\be_1,\mu}|^2\sum_{\mu'}|h_{-\bR,\mu'}|^2
\]
\[
=\frac{4}{\eta^4}\cases{
1 & $\bR=\pm\be_{1,2}$ \cr
1/\eta^2 & $\br=\pm(\be_1+\be_2),\pm(\be_1-\be_2),\pm2\be_{1,2}$  \cr
}
\eeq
and for the corresponding two step term 
$
\langle K_{\br\br'}K_{\br'\br''}\rangle_\alpha-\langle K_{\br\br'}\rangle_\alpha\langle K_{\br'\br''}\rangle_\alpha
$
of Eq. (\ref{two_steps}) 
\beq
\frac{1}{4}h^{}_{-\be_1,\mu}h^\dagger_{0,\mu}h^{}_{-\bR,\mu}h^\dagger_{\bR+\be_1,\mu}
=\frac{1-8/\eta^2}{\eta^2}\cases{
(-1+8/\eta^2) & $\bR=-\be_1$ \cr
2/\eta^2 & $\bR=\be_1, \pm \be_2$ \cr
-2/\eta^2 & $\bR= -\be_1\pm\be_2, -2\be_1$ \cr
0 & otherwise \cr
}
,
\eeq
which is real and does not depend on the band index $\mu$.
It should be noted that (i) $\bR=-\be_1$ implies $\br''=\br$, which does not contribute
to the determinant or to the adjugate matrix, and
(ii) the two step term vanishes for $\bR=\be_1\pm\be_2$, $2\be_{1,2}$ and $-2\be_2$,
indicating a complete suppression of the transfer to these sites.
This, in contrast to the non-vanishing value at $\bR=-\be_1\pm\be_2$ and $\bR=-2\be_1$,
represents the anisotropy due to the rubber-band effect of the phase averaging.

The two examples in Fig. \ref{fig:graph1}b) can be directly extended to larger random walks
by adding repeatedly either a link with a single-site loop (upper graph) or a two-site loop 
$\langle K_{\br\br'}\rangle_\alpha$ (lower graph). With Eq. (\ref{nn_hopping}) the weight
of the upper graph then changes by a factor of order $1/\eta$, while the lower graph changes
by a factor of order $1/\eta^2$ for each additional link. This indicates that for this example 
the weight of the exponentially decaying upper graph dominates the classical random walk
graph as the number of steps increases. 

\section{Conclusions}
\label{sect:conclusions}

The analysis of the invariant measure has revealed that transport in systems with a chiral
Hamiltonian is linked to quantum diffusion of fermions, where the evolution is described by the 
random transition amplitude $K_{\br\br'}$ with particle conservation $\sum_{\br'}K_{\br\br'}=1$.
This conservation law is reflected by a uniform zero mode. 
Central for the derivation of the transition amplitude is the effective Green's function $u$
defined in Eq. (\ref{eff_gf2}). This can be seen as an approximation of the original product 
of Green's functions $G(\pm \epsilon)$. The fluctuations of $u$ are bounded, in contrast to the
unbounded fluctuations of $G(\pm \epsilon)$.
Through phase averaging, i.e., integration with respect to the invariant measure,
two types of competing walks are created, namely classical random walks and back-folded random walks.
The latter are formed as if the walker is pulled back by rubber bands to its previous positions.
The competition of these contributions can lead to a suppression of classical diffusion 
due to interference effects in phase averaging. Details of this phenomenon for specific models
are still open and should be addressed in the future.

The limit $\eta\to0$ is connected with a transition from the regime of quantum diffusion to
a non-diffusive regime. The latter is characterized by a vanishing average density of states. 
Assuming a gap in the spectrum of ${\bar H}$, disorder in the form of, e.g., a random gap 
term $m$ in $H=h_1\sigma_1+h_2\sigma_2 +m\sigma_3$, can create localized (bound) states
at zero energy (midgap states), which are not visible in the average density of states.
The properties of the transition and the transport properties in the regime with $\eta=0$,
where the invariant measure approach is not valid any more, should be studied in a separate project.

\appendix

\section{Separation of the uniform zero mode}
\label{app:zero_mode1}

The role of the uniform zero mode can be discussed in term of the spectral representation of the (normalized) 
correlator before averaging
\[
C^{-1}_{\bR\bR'}
=\sum_{k\ge 0}\frac{\psi^*_k(\bR)\psi_k(\bR')}{\lambda_k}
\ \ (\lambda_0=O(\epsilon))
\] 
with eigenmodes $\psi_k$ and eigenvalues $\lambda_k$ of $C$. Since the uniform zero mode $\psi_0=1/\sqrt{|\Lambda|}$ 
is not random and $\lambda_k\ne0$ for $k>0$, we separate this mode and split the sum into $k=0$ and $k>0$:
\beq
C^{-1}_{\bR\bR'}=\frac{1}{|\Lambda|}{\bar K}_{rec}+
\sum_{k>0}\frac{\psi^*_k(\bR)\psi_k(\bR')}{\lambda_k}
\ ,\ \ 
\frac{1}{|\Lambda|}{\bar K}_{rec}=\frac{\psi^*_0(\bR)\psi_0(\bR')}{\lambda_0}=\frac{1}{|\Lambda|\lambda_0}
,
\eeq
where the recurrent part with ${\bar K}_{rec}$ does not decay but vanishes uniformly for a large system size $|\Lambda|$. 
The recurrent part provides the $1/\epsilon$ behavior for the normalized $\sum_{\bR'}{\hat K}_{\bR\bR'}$, while 
the remaining sum is independent of $\lambda_0$. 
After phase averaging we get for the right-hand side of Eq. (\ref{corr0})
\beq
\frac{\langle C^{-1}_{\bR\bR'}\prod_{k'\ge0}\lambda_{k'}\rangle}{\langle \prod_{k'\ge0}\lambda_{k'}\rangle}
=\frac{\lambda_0\langle C^{-1}_{\bR\bR'}\prod_{k'>0}\lambda_{k'}\rangle}{\lambda_0\langle \prod_{k'>0}
\lambda_{k'}\rangle}
=\frac{1}{|\Lambda|}{\bar K}_{rec}+\frac{1}{\langle \prod_{k'>0}\lambda_{k'}\rangle}
\Big\langle\sum_{k>0}\frac{\psi^*_k(\bR)\psi_k(\bR')\prod_{k'>0}
\lambda_{k'}}{\lambda_k}\Big\rangle
,
\eeq
since $\lambda_0$ does not depend on the random variables. This means that the uniform zero mode gives an
additive term to the correlator, which leads to a divergence of  $\sum_{\bR'}{\hat K}_{\bR\bR'}\sim 1/\lambda_0$ 
even for a finite system.


\end{document}